\documentclass[11pt,a4paper]{article}
\pdfoutput=1

\usepackage{jheppub}

\usepackage{booktabs}
\usepackage{array}
\usepackage{amsmath}
\usepackage{amssymb}
\usepackage{graphicx}
\usepackage{float}
\usepackage{braket}
\usepackage{hyperref}
\usepackage{slashed}
\usepackage{enumerate}

\usepackage[all]{hypcap}

\usepackage{feynmp}
\def\beq{\begin{gather}}
\def\eeq{\end{gather}}
\def\beeq{\begin{eqnarray}}
\def\eeeq{\end{eqnarray}}

\def\gp2{g^{\prime 2}}

\newcommand{\custh}{SU(2)^2\rtimes \mathbb{Z}_2}

\newcommand{\Gev}{\ensuremath{\, \mathrm{GeV}}}
\newcommand{\TeV}{\ensuremath{\, \mathrm{TeV}}}
\newcommand{\GeV}{\ensuremath{\, \mathrm{GeV}}}

\newcommand{\lag}{\mathcal{L}}

\newcommand{\pT}{\ensuremath{p_{T}} }
\newcommand{\met}{\ensuremath{E_{T}^{miss}} }
\newcommand{\ttbar}{\ensuremath{t\overline{t}} }
\newcommand{\meff}{\ensuremath{m_{\textrm{Eff}}} }
\newcommand{\tprime}{\ensuremath{T} }
\newcommand{\bprime}{\ensuremath{B} }
\newcommand{\xfourthird}{\ensuremath{X_{-\frac{4}{3}}} }
\newcommand{\ifb}{\ensuremath{\textrm{fb}^{-1}} }
\newcommand{\mreco}{\ensuremath{m_{\ell \nu b}} }
\newcommand{\mreconospace}{\ensuremath{m_{\ell \nu b}}}

\usepackage{color}

\title{Search Strategies for Top Partners in Composite Higgs models} 
\date{\today}
\author[a]{Ben Gripaios,} 
\author[a]{Thibaut M\"{u}ller,} 
\author[a]{M. A. Parker} 
\author[a]{and Dave Sutherland} 

\affiliation[a]{Cavendish Laboratory, \\ J.J.\ Thomson Avenue, Cambridge, UK}

\emailAdd{gripaios@hep.phy.cam.ac.uk}
\emailAdd{mueller@hep.phy.cam.ac.uk}
\emailAdd{parker@hep.phy.cam.ac.uk}
\emailAdd{dws28@cam.ac.uk}

\preprint{Cavendish-HEP-14/03}

\abstract{
We consider how best to search for top partners in generic composite
Higgs models. We begin by
classifying the possible group representations carried by top partners in
models with and without a custodial $SU(2)\times SU(2) \rtimes
\mathbb{Z}_2$ symmetry protecting the rate for $Z
\rightarrow b\overline{b}$ decays. We identify a number of minimal
models whose top partners only have electric charges of
$\frac{1}{3},  \frac{2}{3},$ or $\frac{4}{3}$ and thus decay to top or
bottom quarks via a single Higgs or electroweak gauge boson.  We
develop an inclusive search for these based on a top veto, which we find to be more
effective than existing searches.
Less minimal models feature light states that can be sought in
final states with like-sign leptons and so
we find that 2 straightforward LHC searches give a reasonable
coverage of the gamut of composite Higgs models.
}  

\begin{document}   
 
\maketitle 

\section{Introduction}
\label{sec:intro}
Models in which the naturalness problem of the Standard Model (SM) is solved by making the Higgs
a composite, pseudo-Nambu-Goldstone boson (PNGB) of some
as-yet-unknown strong
dynamics \cite{Kaplan:1983fs,Georgi:1984af,Dugan:1984hq} have come to the fore in recent years as they have
matured
\cite{Contino:2003ve,Agashe:2004rs,Giudice:2007fh},
incorporating symmetries \cite{Sikivie:1980hm,Agashe:2006at} and partial compositeness \cite{Kaplan:1991dc} to mitigate
unobserved corrections in electroweak precision tests
and flavour physics. They are, by now, arguably just as good (or just as bad) a
contender as supersymmetry for new physics at the TeV scale,
and the LHC experiments ought to devote significant resources to looking
for them.

Just as for supersymmetry, the composite Higgs is more a
paradigm than a model, in that it may be realised in arbitrarily many
ways. Thus it is not clear, {\em a priori}, how best to search at the LHC for
evidence of a composite Higgs. Na\"{\i}vely, the obvious place to look is in the Higgs
sector itself, but the minimal model \cite{Agashe:2004rs} contains
just a single SM Higgs doublet (for models with an extended Higgs
sector, see, {\em e.g.} \cite{Gripaios:2009pe,Mrazek:2011iu}).
Moreover, the apparent absence of new physics in
the electroweak and flavour sectors already forces us into a
slightly-tuned regime in which
the electroweak scale is suppressed compared to the scale of strong
dynamics, which suggests 
that the couplings of a composite Higgs to
other SM particles will deviate by only a few percent from the SM
values \cite{Falkowski:2007hz,Low:2009di}.

The necessary suppression also implies that new states, beyond the SM,
will generically be out of the mass reach of the LHC. However, the mechanism of partial compositeness requires the
existence of additional states that mix with the top quark and several authors
have argued \cite{Contino:2006qr,Matsedonskyi:2012ym,Redi:2012ha,Marzocca:2012zn,Pomarol:2012qf,Panico:2012uw,Pappadopulo:2013vca} that these ``top partners'' should be light, given the measured
Higgs boson mass. In a nutshell, the connection arises because the
Higgs potential (and, {\em ergo}, the Higgs mass) is predominantly induced by radiative corrections
involving the top quark, which are cut off
by the top partners.

Many suggestions for dedicated searches for such top
partners have been put forward \cite{Contino:2008hi,deSandes:2008yx,AguilarSaavedra:2009es,Mrazek:2009yu,Dissertori:2010ug,Gopalakrishna:2011ef,Harigaya:2012ir,Vignaroli:2012nf,topPartnerHuntersGuide,Gopalakrishna:2013hua,Banfi:2013yoa,Azatov:2013hya,Gillioz:2013pba,Matsedonskyi:2014lla,Beauceron:2014ila,Ortiz:2014iza,Barducci:2014ila,Endo:2014bsa}. Indeed, there are already more
on the market than the experiments have the resources to implement,
with just a few analyses having appeared thus far. 

We thus consider the question of whether the experiments could obtain
a reasonable coverage of the space of composite Higgs models with
just a small number of generic searches. To do so, we must first classify
the possible top partners and their phenomenological implications. In all the models that have
appeared so far, the top partners
are colour triplets, just like the top quark,\footnote{Models in which
  the lightest
  top partners are not colour triplets are conceivable, but will
  typically feature leptoquark states \cite{Gripaios:2009dq}, which themselves
  provide an alternative target for
  LHC searches \cite{Gripaios:2010hv}.} and so can be pair-produced at the LHC via
strong interactions with significant cross-section.\footnote{There may
  also be a significant cross-section for single production, in association
  with quarks.} It is the subsequent decays that give us the most 
phenomenological room to man\oe uvre. The possible decays are, of course, restricted
by the electric charge, which must equal $\pm \frac{2}{3}$ modulo any
integer, and we shall see below that any such charge is possible for
the lightest top partner.

If the lightest states have charge not equal to
$\frac{1}{3}$, $\frac{2}{3}$, or $\frac{4}{3}$, then they
decay via multiple $W$ bosons and lead to easily identifiable final
states with like-sign leptons \cite{Contino:2008hi}. Other states pose more of
a problem for experimental searches, since they can decay to
a variety of states ($Wq,Zq,$ or $Hq$,
where $q$ is a top or bottom quark) with comparable branching
fractions at most points in parameter space.
 We find that, rather than
developing specific searches targeted towards the different decay channels, it is
more effective to simply reduce the dominant SM backgrounds
(especially $t\overline{t}$) by vetoing events with a top quark, in a
certain way.

The outline is as follows.
In the next Section, we begin with a group-theoretic classification of
possible top
partners. We show that in general it is possible to have a lightest
top partner with any electric charge, but that there are restrictions
if we ask that the theory have a symmetry protecting the decay rate for $Z\rightarrow b
\overline{b}$ \cite{Agashe:2006at}. Restrictions also arise if we
insist that the right-handed top, $t_R$, be wholly
composite. The results will, no doubt, come as no surprise to experts, but we have not been able to find them in the literature.
In \S\ref{sec:models}, we discuss the minimal models that only
contain top partners with charge
$\frac{1}{3}$, $\frac{2}{3}$, or $\frac{4}{3}$. These have top partners in 1-, 2-, and 3-
dimensional representations, and there are 2 inequivalent
representations of each dimension, giving 6 models in all. The 1-d models (discussed already in \cite{topPartnerHuntersGuide}) allow for both
custodial protection of $Z\rightarrow b
\overline{b}$ and $t_R$ compositeness. The other models allow neither,
but are of interest as straw models, as their
phenomenology is rather different. In the
doublet and triplet models, the lightest states have charges
$\frac{1}{3}$ and $\frac{4}{3}$, respectively. We also give an
explicit 
effective Lagrangian description for the doublet models, following the
approach pioneered for the 1-d models in \cite{topPartnerHuntersGuide}.
In \S\S\ref{sec:currentLimits}-\ref{sec:newanalysis}, we estimate the limits on these models
that are obtained using a slew of existing LHC searches and show how a
single search based on a top veto is more effective in all cases.
\section{Taxonomy of Top Partners}
\label{sec:group}
We first wish to classify the possible
top partners, which requires us to discuss their representations
(henceforth `reps') and
in turn the groups whose reps they carry. 
Since we wish to include in our discussion the custodial symmetry
for $Z \rightarrow b\overline{b}$, it will not suffice to discuss Lie
algebras; we must discuss Lie groups. We give only the salient
definitions and results here, relegating most of the details
and proofs to Appendix \ref{app:lie}.

A composite Higgs model is based on a coset $G/H$, and the usual rules
of non-linear sigma models \cite{Coleman:1969sm} dictate that the top partner
must carry a rep of $H$. As we argue in Appendix
\ref{app:lie}, it suffices to consider $G = Sp(2)$ and either $H =
SU(2)\times SU(2)$ (henceforth denoted $SU(2)^2$)
or $H = \custh $, depending on whether we wish to have custody
of $Z \rightarrow b\overline{b}$ or not. In the semi-direct
product in $\custh$, the $\mathbb{Z}_2$ is mapped to the
automorphisms of $SU(2)^2$ given by permutations of the 2
$SU(2)$ subgroups.

The unitary irreducible reps (irreps)
of $Sp(2)$ can be labelled by $(m,k)$, where $m \geq k \geq 0$ and $2m+1$
and $2k+1$ are either both even or both odd, and have dimension $ (2k + 1)(2m + 3)(m +
k + 2)(m- k + 1)/6$. The irreps of $SU(2)^2$ can be labelled by
$(2l+1,2r+1)$, where $2l+1 \in \mathbb{N}$ denotes the dimension of the spin-$l$
irrep of $SU(2)$ (and similarly for $r$), and have dimension $(2l+1)(2r+1)$. The irreps of
$\custh$ fall into 3 classes, which we label by $(2l+1,2l+1)^\pm$ and
$((2l+1, 2r+1))$, of dimensions $(2l+1)^2$ and $2 (2l+1)(2r+1)$, respectively. The
branching rules for $Sp(2) \rightarrow \custh $ are given in Appendix
\ref{app:lie}, and are shown explicitly in Table \ref{tab:reps} for dimensions up
to 50.  The
branching rules for $\custh \rightarrow SU(2)^2 $ are $(2l+1,2l+1)^\pm
\rightarrow (2l+1,2l+1)$ and $((2l+1, 2r+1)) \rightarrow (2l+1, 2r+1)
\oplus (2r+1, 2l+1)$.

The classification of top partners is then as follows.
If we desire custody of $Z \rightarrow
b\overline{b}$, the top partner must carry one of the irreps $(2l+1,2l+1)^\pm$
or the irrep $((2l+1,2l+3))$, for some $l$.  If we desire a fully composite $t_R$, then the top partner must carry one of
the irreps $(2l+1,2l+1)^\pm$ of $\custh$ (or $(2l+1,2l+1)$ of
$SU(2)^2$). If we desire neither, the top partner may carry any irrep
of $SU(2)^2$. 

For each of these top partner reps, we are still left with an infinity
of possible models, because we are free to assign the
elementary $Q_L$ to a spurionic irrep of $Sp(2)$ with arbitrarily large
$2m+1$. 
For example, in a model with a top partner carrying either the
$(1,1)^+$ or $(1,1)^-$, respectively, of
$\custh$,
the $Q_L$ can be assigned to any irrep $(m,0)$ of $Sp(2)$ with $m$ even or
odd, respectively.  

Models with top partners carrying either the 
$(1,1)$ (or $(1,1)^\pm$), $(2,1)$ (or $(1,2)$), or $(3,1)$ (or $(1,3)$) irreps pose a
particular challenge phenomenologically, since these
can feature only states with charges
$\frac{1}{3}$, $\frac{2}{3}$, or $\frac{4}{3}$. In what follows, we
consider the prospects for experimental searches in these worst-case
scenarios. To get a reasonable idea of the limits
that can be obtained, it suffices to consider just one model for each
possible dimension of the top partner irrep. Other models (including those
with different choices of the $Sp(2)$ irrep) have the same spectrum
of states, but with varying branching ratios. Since our proposed
search includes all decay modes, the limits obtained should not vary
too significantly.

\begin{table}
\begin{tabular}{c  c  c  }
\hline
$Sp(2)$ irrep $(m,k)$ & dimension & $\custh $ irreps \\ \hline
$(0, 0)$ & 1 & $(1, 1)^+$ \\
 $(\frac{1}{2}, \frac{1}{2})$ & 4 & $((2, 1))$ \\
 $(1, 0)$ & 5 & $(1, 1)^- \oplus (2, 2)^-$ \\
$ (1, 1)$ & 10 &$((3, 1)) \oplus (2, 2)^+$
\\ $(2, 0)$& 14& $(1, 1)^+ \oplus (2, 2)^+ \oplus (3, 3)^+$ \\
$ (\frac{3}{2}, \frac{1}{2}) $& 16 & $((2, 1)) \oplus  ((3, 2)) $\\
$(\frac{3}{2}, \frac{3}{2})$ & 20 & $((4, 1)) \oplus ((3, 2))$ \\
$(3, 0) $& 30& $(1, 1)^- \oplus  (2, 2)^- \oplus  (3, 3)^- \oplus (4,
4)^- $\\
$(2, 2)$ & 35 & $((5, 1)) \oplus ((4, 2)) \oplus (3, 3)^+$ \\
$ (2, 1)$ & 35&$ ((3, 1)) \oplus (2, 2)^- \oplus  ((4, 2)) \oplus (3, 3)^-$ \\
$ (\frac{5}{2}, \frac{1}{2})$ & 40 &$((2, 1)) \oplus ((3, 2)) \oplus
((4, 3))$ \\ \hline
\end{tabular}
\caption{ \ensuremath{Sp(2) \rightarrow \custh} branching rules for dimensions
  up to 50. \label{tab:reps}}
\end{table}

\section{Phenomenology of the singlet, doublet and triplet models}
\label{sec:models}
In what follows, we will only discuss experimental searches for 3
of the minimal models just described, namely a $(1,1)^-$ model containing only a \tprime (charge $\frac{2}{3}$); a $(1,2)$ model
containing a \tprime and a \bprime (charge
$-\frac{1}{3}$), and a $(3,1)$ model containing a \tprime, \bprime and \xfourthird (charge
$-\frac{4}{3}$).\footnote{Note that the triplet charge assignments are
  incompatible with custody of $Z\rightarrow b\overline{b}$, which
  requires charges of $-\frac{1}{3}, \frac{2}{3},$ and $\frac{5}{3}$
  for the $((1,3))$ irrep.} None of these states have exotic-looking
decays, and thus these models present the most challenging
experimental signatures. Higher-dimension representations
necessarily include states with charge not equal to
$\frac{1}{3}$, $\frac{2}{3}$, or $\frac{4}{3}$, decaying via multiple
$W$ bosons to like-sign leptons \cite{Contino:2008hi}. This is a more straightforward experimental
signature, which is already being explored at the LHC~\cite{CMSSameSignTopPartner, ATLASSameSignTopPartner,Matsedonskyi:2014lla} (see \S\ref{sec:nonmin}).

\subsection{Singlet model}
Our first model, already described in
\cite{topPartnerHuntersGuide} (where it is called `$\mathrm{M}1_5$') has
the advantages of featuring custody of $Z\rightarrow
b\overline{b}$ and full $t_R$ compositeness. The model has the
following fields:
$Q_L$, carrying the 5-d $(1,0)$ irrep of $Sp(2)$ and containing the
elementary $t_L$ and $b_L$; $\Psi$, carrying the 1-d $(1,1)^-$ irrep of
$\custh$ and containing one vector-like top partner $T$; $t_R$, also carrying the $(1,1)^-$ irrep of
$\custh$ and containing the composite $t_R$;\footnote{Desiring to
  avoid overloading the reader's RAM, and at the risk of incurring
  kernel panic, we use $t_R$ to
  represent both the Weyl fermion and the vector that gives its
  embedding in the 5 rep of $Sp(2)$: $t_R = \left( 0, 0, 0, 0, t_R
  \right)^T$.}\footnote{As in \cite{topPartnerHuntersGuide}, the $b_R$
  is absent from our discussion.} and the coset representative, $U$, built out of the four real Higgs fields (the exact embeddings are given in \cite{topPartnerHuntersGuide}). The effective Lagrangian is
\begin{align}
 \lag &= i \overline{Q_L} \slashed{D} Q_L + i \overline{t_R} \slashed{D} t_R + \overline{\Psi} (i \slashed{D} - M) \Psi \nonumber \\
 & -y f \overline{Q_L} U \Psi_R - y c_2 f \overline{Q_L} U t_R +
 \mathrm{h.c} +\dots,
\label{eqn:singletlag}
\end{align}
parametrised by $\{f,M,y,c_2\}$, of which $c_2$ is fixed to obtain the correct top mass as follows. After EWSB, the charge $\frac{2}{3}$ states mix according to the matrix
\begin{equation}
 \lag \supset - \left( \begin{matrix} \overline{t_L} & \overline{T_L} \end{matrix} \right) \left( \begin{matrix} \frac{y c_2 v}{\sqrt{2}} & \frac{y v}{\sqrt{2}} \\ 0 & M \end{matrix} \right) \left( \begin{matrix} t_R \\ T_R \end{matrix} \right) .
\end{equation}
where $v=246 \Gev$. The mass matrix may be diagonalised by a singular
value decomposition, where the physical masses are given by the
singular values. We require the smaller singular value to be the top
mass, which implicitly fixes $c_2$ as a function of $M$ and $y$; the
larger value --- now a function of $M$ and $y$ alone --- gives the mass of the top partner.

The top partners are pair produced by QCD interactions, and hence the
pair production cross-section at leading order depends only on the top
partner mass. They may also be singly produced, with LO cross-section
proportional to $y^2$. It is found that the top partner production
cross-section does not strongly depend on the value of $f$ (which we
set to 500~\GeV). We thus consider the free parameters of the model to
be $m_{\tprime}$ and $y$. 
 
There are 3 possible decay modes for the \tprime: $Wb$, $Zt$ or
$Ht$; their branching fractions are roughly 2:1:1.\footnote{This is
  understood~\cite{topPartnerHuntersGuide} by considering the
  interaction $y f \overline{Q_L} U \Psi_R$ of Lagrangian
  (\ref{eqn:singletlag}). In the limit of a heavy $\Psi$, the
  Goldstone boson equivalence theorem predicts equal decays to
  $W^1,W^2,Z$ and $h$.} The final state of singly or pair-produced
\tprime of mass of a few hundred GeV thus invariably contains several high \pT objects, including several $b$-jets, as well as a reasonable branching fraction to leptons. 

\subsection{Doublet model}

The next model contains a top partner carrying the $(1,2)$ irrep of
$SU(2)^2$. Unlike the singlet model, this cannot feature custody of $Z\rightarrow
b\overline{b}$; nor can the $t_R$ be fully composite. 
Neither can be considered a {\em sine qua non}, given the
tunings already required in the electroweak $S$ parameter and flavour
sector and moreover this model 
provides a sufficiently different
experimental signature to merit consideration, especially if it can be
covered by the same experimental search as the singlet model. Summarising
Appendix~\ref{app:model}, this requires elementary fields
$Q_L$, $t_R$ and $b_R$, each carrying the $(\frac{1}{2},\frac{1}{2})$
rep of $Sp(2)$ and containing respectively the $t_L$ and $b_L$, the
$t_R$, and the $b_R$. The top partner field, $\Psi$, contains a $T$
and a $B$. The Lagrangian is
\begin{align}
 \mathcal{L} &= i \overline{Q_L} \slashed{D} Q_L + i \overline{t_R} \slashed{D} t_R + i \overline{b_R} \slashed{D} b_R + \overline{\Psi} (i \slashed{D} - \slashed{e} - M) \Psi \nonumber \\
 & -y f \overline{Q_L} U \Psi_R - y_R^t f \overline{t_R} U \Psi_L -
 y_R^b f \overline{b_R} U \Psi_L + \mathrm{h.c} +\dots.
\end{align}
The model has parameters $\{f,M,y,y^t_R,y^b_R\}$; the latter two of these are fixed by the top and bottom masses after singular value decomposition of the mass matrices.

Again, the phenomenology does not depend strongly on the value of $f$,
which is set to be 500~\GeV, so the free parameters may again be taken
to be $M$ and $y$. For large $y$, the masses of the \tprime and \bprime
are very similar in value, whilst the \tprime is significantly heavier
at small $y$. The \tprime decay channels are as in the singlet model, plus possibly a small branching fraction for $\tprime \rightarrow BW$: the branching fraction to $\tprime \rightarrow ht$ dominates ($\sim 50 - 80\%$ in the parameter space considered). The \bprime has three decay modes: $\bprime \rightarrow Wt$, $\bprime \rightarrow Zb$, $\bprime \rightarrow hb$. The decay via a $W$ boson is the most likely, with a branching fraction of $\sim 50 - 80\%$; the other two decays are equally likely. The phenomenology is thus more varied than for the singlet model, but the most distinctive features are still present: large $b$-jet multiplicity and some likelihood of leptonic $W$-boson decays at high mass.

\subsection{Triplet models}
As described above, there are also triplet models with top partners
whose charges only equal
$\frac{1}{3}$, $\frac{2}{3}$, or $\frac{4}{3}$, {\em viz.} $(1,3)$ or
$(3,1)$ irreps. The novel feature
of these models is that their lightest top partner, in the absence of similarly charged states to mix with, has charge $\pm
\frac{4}{3}$, and will overwhelmingly decay to $b$ quarks via `wrong-sign'
$W$-bosons. Excepting the unmeasured charge of the b quark, this
is the same $Wb$ signature which is targeted in the case of the
singlet model. Compared to the singlet case, however, the signal is
enhanced by virtue of $Br(\rightarrow Wb) = 1$ and by the additional
contribution of the triplet's charge $\frac{2}{3}$ state. The same
search strategy will thus give limits that are at least as good as the singlet model
for corresponding masses, and we do not consider them further here.

\subsection{Larger representation models}
\label{sec:nonmin}
As all less minimal models contain states with charges not equal to
$\frac{1}{3}$, $\frac{2}{3}$, or $\frac{4}{3}$
(such as $\frac{5}{3}$, \dots), they present a more straightforward
experimental signature, in the form of same-sign leptons, and searches
are already being performed by the ATLAS and CMS
Collaborations~\cite{CMSSameSignTopPartner, ATLASSameSignTopPartner}.

Not only are these states easily identifiable, but they are also the
lightest, as we now prove. In our effective field theory approach, we
integrate out all but the top partner $\Psi$, {\em i.e.}\ the lightest
fermionic resonance of the strong sector. For (a) a partially
composite and (b) a fully composite $t_R$, for example, the mass terms in the Lagrangian are:
\begin{enumerate}[(a)]
 \item $\lag \supset -M \overline{\Psi_L} \Psi_R - y_1 f \overline{Q_L} U \Psi_R - y_2 f \overline{Q_L} U t_R +\text{h.c.} $,
 \item $\lag \supset -M \overline{\Psi_L} \Psi_R - y_1 f \overline{Q_L} U \Psi_R - y_2 f \overline{t_R} U \Psi_L +\text{h.c.}$.
\end{enumerate}

The mass matrices for the charge $\frac{2}{3}$ states are thus
\begin{enumerate}[(a)]
 \item $\left( \begin{matrix} t_L & T^1_L & T^2_L & \ldots & T^n_L \end{matrix} \right) \left( \begin{matrix} a & a_1 & a_2 & \ldots & a_n \\ 0 & M & 0 & & \\ 0 & 0 & M & & \\ \vdots & & & \ddots & \\ 0 & & & & M \end{matrix} \right) \left( \begin{matrix} t_R \\ T^1_R \\ T^2_R \\ \vdots \\ T^n_R \end{matrix} \right) $,
 \item $\left( \begin{matrix} t_L & T^1_L & T^2_L & \ldots & T^n_L \end{matrix} \right) \left( \begin{matrix} 0 & a_1 & a_2 & \ldots & a_n \\ b_1 & M & 0 & & \\ b_2 & 0 & M & & \\ \vdots &  & & \ddots & \\ b_n & & & & M \end{matrix} \right) \left( \begin{matrix} t_R \\ T^1_R \\ T^2_R \\ \vdots \\ T^n_R \end{matrix} \right)$,
\end{enumerate}
whose singular values under singular value decomposition give the
fermion masses. $a_i \sim \mathrm{O}(y_1 v)$ and $b_i \sim \mathrm{O}(y_2 v)$ with $v$ the SM Higgs v.e.v., and assuming $a_i,b_i \ll M$ then in both cases (a) and (b) the resulting spectrum of masses
is $M$ (with degeneracy $n-1$) together with 1 value $\leq M$ (which
we identify with $m_t$) and one
value $\geq M$. The same is true for the charge $-\frac{1}{3}$
states. This can be seen explicitly for case (b), for example, as follows. By suitable rotations of the left and right handed composite states the mass matrix (b) may be put in the form:
\begin{equation*}
 \left( \begin{matrix} 0 & a & 0 & \ldots & 0 \\ b & M & 0 & & \\ 0 & 0 & M & & \\ \vdots &  & & \ddots & \\ 0 & & & & M \end{matrix} \right),
\end{equation*}
which has $n-1$ singular values equal to $M$, and the other two are the singular values of $\left( \begin{matrix} 0 & a \\ b & M \end{matrix} \right)$, namely
\begin{equation*}
 \sqrt{ \frac{1}{2}(M^2 + a^2 + b^2) \pm \frac{1}{2} \sqrt{(M^2 + a^2 + b^2)^2 - 4 a^2 b^2}} \approx \begin{cases} \sqrt{M^2 + a^2 + b^2}, & \text{top partner}, \\ \frac{ab}{M}, & \text{top}. \end{cases}
\end{equation*}
These solutions are, respectively, $>M$ and $<M$. Note also that, in
the case where the model contains a charge $-\frac{1}{3}$ $B$ as well
as a charge $\frac{2}{3}$ $T$ top partner, the values $a_i$ and $M$
are common to both charge $-\frac{1}{3}$ and $\frac{2}{3}$ mass
matrices. The $b_i$ are adjusted to produce a bottom mass smaller than
that of the top, and in doing so produce a $B$ lighter than the $T$. A similar proof holds for case (a).

In contrast to the charge $-\frac{1}{3}$ and $\frac{2}{3}$ composite states, the other
states have nothing to mix with, and have mass $M$, which is the
lightest top partner mass.


\section{Current experimental limits on the singlet and doublet models}
\label{sec:currentLimits}
Current experimental limits can be divided into 2 categories: those from
dedicated searches for $\tprime$ and $\bprime$, and those from other
analyses searching for other models, but considering similar
final-state signatures. Particularly relevant are SUSY searches
requiring high effective mass, high $b$-jet multiplicity, and leptons. We find
that the limits from the dedicated searches are generally the
strongest, as might be expected, but that at large coupling $y$, the
SUSY searches become competitive. We now describe the searches and our
reinterpretation of their limits. The results are shown in \S\ref{sec:limits}.\footnote{Considering just pair production, model-independent limits on top partners from both SUSY and dedicated searches can be found in \cite{Barducci:2014ila} (in terms of their masses and branching fractions).}

\subsection{SUSY searches}
As most SUSY searches are ``cut-and-count", they can be readily reinterpreted in the context of composite Higgs models through the raw event count in the signal regions. For each model, a grid of signal points is simulated using FeynRules~\cite{Alloul:2013bka} and {\sc MadGraph}~5 v1.5.12~\cite{Maltoni:2002qb}. The hadronic shower was simulated using {\sc Pythia}~6~\cite{Sjostrand:2006za}, and the response of the ATLAS detector was simulated using {\sc Delphes}~3~\cite{deFavereau:2013fsa}. The same cuts as for the searches are applied, and limits are then set by comparing the observed and expected number of events from the published search with the signal expectation, using the  $CL_s$ prescription~\cite{clsTechnique}.

Of all the public searches for SUSY by the ATLAS and CMS Collaborations that we tested, 3 were found to be reasonably sensitive. They are the ATLAS ``0 $\ell$ + 2 $b$-jets + $\met$"~\cite{ATLAS0Lepton2BJets}, the ATLAS ``1 $\ell$ + 2-6 jets"~\cite{ATLASOneLeptonJets} and the CMS ``1 $\ell$ + 6-7 jets + $b$-jets"~\cite{CMS1Lepton68Jets15bJets} searches.

\subsection{Dedicated searches}
Both the ATLAS and CMS Collaboration have performed dedicated searches
for $\tprime$ and $\bprime$ top partners~\cite{CMSTopPartnerSingleLepton, ATLASTopPartner8TeVCONF, CMSBSearch}. These searches only target the pair-production of the new states in one or two very specific decay channels, and set limits on their masses in terms of their branching fractions. These limits can then be used directly to set limits on the models considered here. These searches make use of advanced techniques that are not easily replicable, such as BDTs, jet substructure and hadronic $W$-tagging; we believe these results can be reproduced more simply and effectively using the search in section \S\ref{sec:limits}.
\section{New analysis design}
\label{sec:newanalysis}
Expanding upon some of the search strategies already implemented by
the LHC experiments, we show that it is possible to set stringent
limits on both the singlet and the doublet model with one rather straightforward and inclusive search. We estimate limits on the current models from this search for both the current LHC dataset (20.3 \ifb at $\sqrt{s} = 8~$TeV), as well as after a year of running at higher energies (30 \ifb at  $\sqrt{s} = 14~$TeV). All signal samples are simulated using {\sc MadGraph}~5 v1.5.12, showered with {\sc Pythia}~6, and the ATLAS detector response is simulated using {\sc Delphes}~3. Jets are reconstructed using the anti-$k_T$ algorithm with $R = 0.5$.

For the $\sqrt{s} = 8~$TeV analysis, events are required to contain
exactly 1 lepton, at least 1 $b$-jet and at least 5 jets in
total. Furthermore, events must have $\met > 50~\GeV$, the leading jet
must have $\pT > 400~\GeV$ and $\meff > 1300~\GeV$.\footnote{The
  scalar sum of the transverse momenta of all selected jets and
  leptons as well as the $\met$.} The $\sqrt{s} = 14~\TeV$ analysis differs from this only in that the jet \pT cut is raised to 600~\GeV, and the \meff cut to 1500~\GeV.
In addition to these cuts, both analyses use a ``top veto'', requiring that the minimum reconstructed mass of the decay $X \rightarrow W b \rightarrow \ell \nu b$, \mreco, be above $200~\GeV$. The analysis cuts are summarised in Table~\ref{tab:analysisSelection}.

The top veto, being the most unusual feature of the analysis,
deserves some further discussion. The invariant mass of the decay is
reconstructed by first solving for the $z$-component of the missing
transverse momentum, using the assumption that the lepton and
missing energy come from the leptonic decay of a $W$ boson (of known
mass $m_W$). To do so requires us to solve a quadratic equation, and
thus there is a twofold ambiguity in the resulting $W$-boson
4-momentum. These two 4-momenta are then added in turn to each possible $b$-jet
4-momentum in the event. The 4-momentum with the minimum invariant
mass, \mreconospace, that results is selected. For illustration, the
distribution of $\mreconospace$ is shown in Figure~\ref{fig:compHRecnstructedMass}. The selection cuts are somewhat loosened with respect to the signal selection to increase statistics: 3 instead of 5 jets, leading jet satisfying $\pT > 300~\GeV$ instead of $\pT > 400~\GeV$, $\meff > 1000~\GeV$ instead of $1300~\GeV$, and no \mreco cut.

\begin{figure}[h!]
	\centering
	\includegraphics[width=0.7\linewidth]{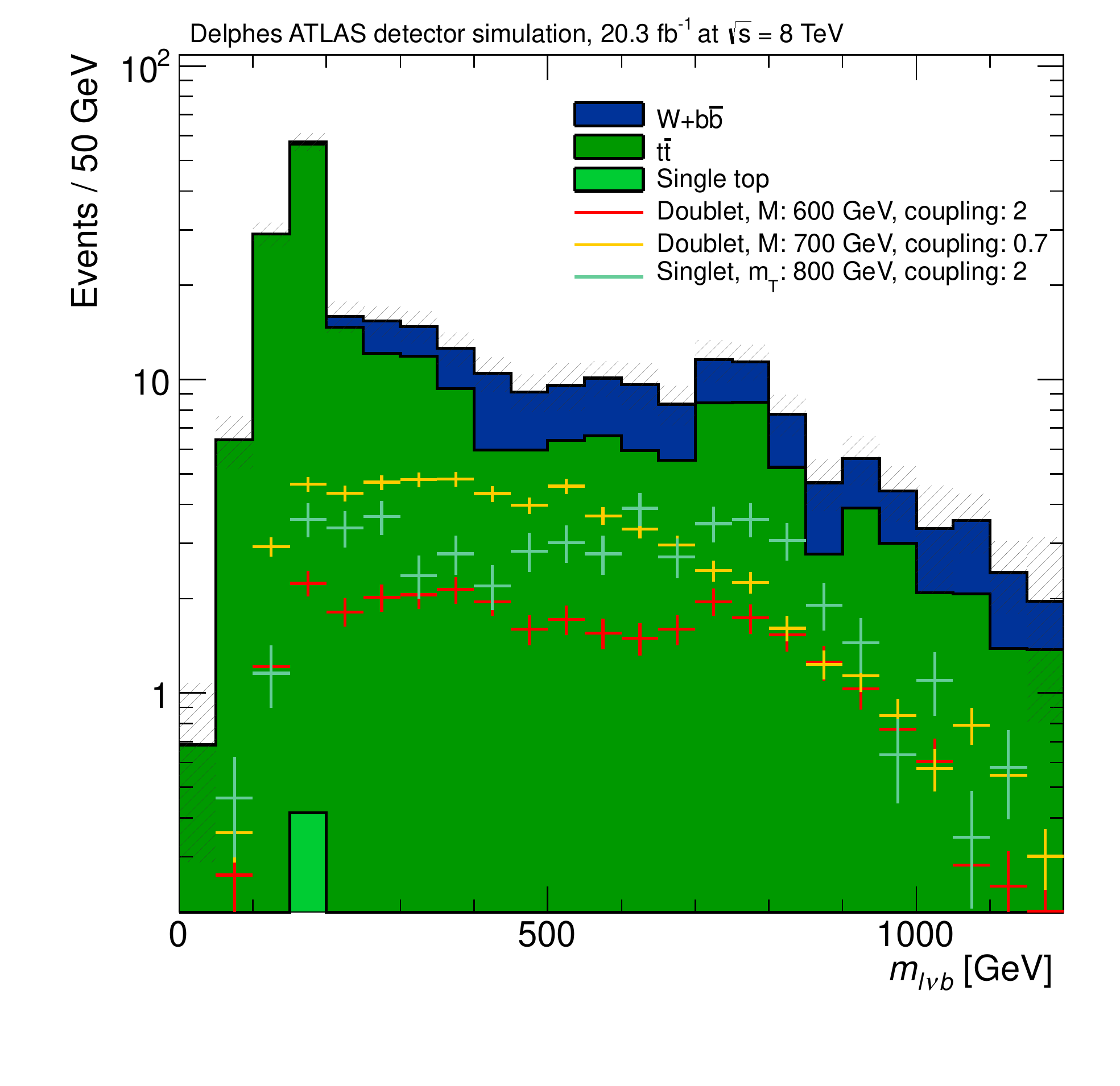}
\caption{Reconstructed mass, $\mreco$, for 20.3 \ifb at $\sqrt{s} =
  8~$TeV, with several superimposed signal distributions for both the
  singlet and doublet models. To increase statistics, the event
  selection for this plot is somewhat loosened from the signal
  selection (for details, see the text).}
\label{fig:compHRecnstructedMass}
\end{figure}

Requiring that this mass lies somewhat above the mass of the top
quark will give a significant rejection of the SM semi-leptonic
\ttbar background. Assuming perfect detector reconstruction, roughly half of the background can be rejected in this way (since there are two
$b$s in the event). The signal suppression is typically rather
less. This is easily seen by considering the possible signal final
states in turn. The worst cases are $B \rightarrow bZ$ and $B
\rightarrow bh$, which contain $0$ or $2$ leptons, and are vetoed (we
require exactly 1 lepton to reduce the SM $Z+$ jets background). But
these tend to have small branching ratio. For $T \rightarrow tZ$,
there can be exactly 1 lepton from the top, but then the probability to
reconstruct a top is comparable to that for the background. 
More promising is pair production of
$B \rightarrow Wt$, where 
 there are twice as many $W$s in the event, compared to the \ttbar background, and so the probability
 that the lepton came from a $t$ quark decay is reduced by a factor of
 one-half. For events involving $T \rightarrow tH$,
 there are extra $b$ jets coming from the Higgs decay that make it
 combinatorially unlikely to reconstruct a $t$. For other decays,
 $\mreco$ will typically be much greater than $m_t$.

Thus we already typically gain a factor of a few in the ratio of
signal to background from the top veto in a given production and decay mode. An even greater gain comes
from the inclusivity of the search that it allows. Whereas existing
searches tune the cuts to focus on a particular production mode
(usually pair production) of a particular state ($T$ or $B$) with a
particular decay mode, our proposed search is sensitive to almost all
of them. Given that the states are generically not too dissimilar in
mass, and given that the branching ratios to different final states
are typically comparable, this leads to rather enhanced sensitivity,
which we expect to see translated into higher limits. 

\begin{table}
	\centering
	\footnotesize
		\begin{tabular}{|l|c|}
			\hline
			\multicolumn{2}{|c|}{\textbf{Object Selection}} \\
			\hline
			\textbf{Object}                                & \textbf{Requirement} \\
			\hline 
			Electrons      &  $\pT > 20$~\GeV, $|\eta| < 2.47$ \\ \hline
			Muons      &   $\pT > 20$~GeV, $|\eta| < 2.4$  \\ \hline
			Jets      &  anti-$k_t$ jets, $R = 0.5$ \\
			&  $\pT > 40$~GeV, $|\eta| < 2.8$  \\ \hline
			Isolation       &  Objects need to be isolated from \\
			& each-other by $\Delta R > 0.5$\\ 
			\hline \hline
			\multicolumn{2}{|c|}{\textbf{Event Selection}} \\
			\hline 
			\multicolumn{2}{|c|}{exactly one lepton} \\ \hline
			\multicolumn{2}{|c|}{number of $b$-jets $\geq 1$} \\ \hline
			\multicolumn{2}{|c|}{$\met > 50$~\GeV} \\ \hline
			\multicolumn{2}{|c|}{number of jets $\geq 5$} \\ \hline
			\multicolumn{2}{|c|}{$\mreco > 200~\GeV$} \\ \hline \hline
			
			\textbf{ $\sqrt{s} = 8~\TeV$ } & \textbf{ $\sqrt{s} = 14~\TeV$ } \\ \hline
			leading jet $\pT > 400$~\GeV  & leading jet $\pT > 600$~\GeV \\ \hline
			$\meff > 1300~\GeV$ & $\meff > 1500~\GeV$ \\ \hline 
			
		\end{tabular}
	\caption{Analysis selection for the 8~\TeV~analysis. The events are passed through the {\sc Delphes} ATLAS detector simulation.}
	\label{tab:analysisSelection}
\end{table}

The main backgrounds to this analysis are \ttbar and $W+b\bar{b}$ (roughly equal proportion), with a small contribution from single top processes, all of which are simulated using the same chain as the signal samples. Contributions from  $Z+$jets processes are vetoed by requiring exactly one lepton, and di-boson contribution is found to be negligible, compatible with the findings of~\cite{ATLASTopPartner8TeVCONF}, which has a similar selection.

\section{Limits on the singlet and doublet models}
\label{sec:limits}
Limits are set using a $CL_s$ test, and assuming that the observed
data would be the same as the expected background.  Statistical
uncertainties are assessed on both the background expectation and the
signal prediction. To estimate the systematic uncertainties on the
background expectation, e.g. from the choice of generator or from
experimental uncertainties, we use the uncertainties on the
backgrounds given in~\cite{ATLASTopPartner8TeVCONF}, where a similar
selection was used. Thus, we assign a 60\% systematic uncertainty on the \ttbar background, and a 42\% uncertainty on the other backgrounds.
The cuts for the analysis were optimised, though the chosen cuts on
some variables such as \meff were not as stringent as they could have
been, as the background estimation on the tails of the distributions
becomes unreliable. In a real experimental search, the background
estimation including fake leptons would be done more carefully and
with more computational resources, and thus a more robust estimate of
the backgrounds would be possible. Thus, the results presented in this
section should be regarded as a rough estimate of the actual limits
that such an
analysis would yield at at $\sqrt{s} = 8~\TeV$ and $\sqrt{s} =
14~\TeV$. The exclusions obtained with the new analysis for the
singlet and doublet models, both at $\sqrt{s} = 8$~$\TeV$ and $\sqrt{s}
= 14$~\TeV, are shown in Figures~\ref{fig:compHAllExclusionsSinglet}
and \ref{fig:compHAllExclusionsDoublet}.

As expected, the use of the top veto and the gain in inclusivity of
the search suggested here yield significantly better limits than
existing searches that target only one decay mode, with dramatically
decreased acceptance. This is especially true for the limits on the
doublet model obtained using dedicated searches.

\begin{figure}[h!]
	\centering
	\includegraphics[width=0.8\linewidth]{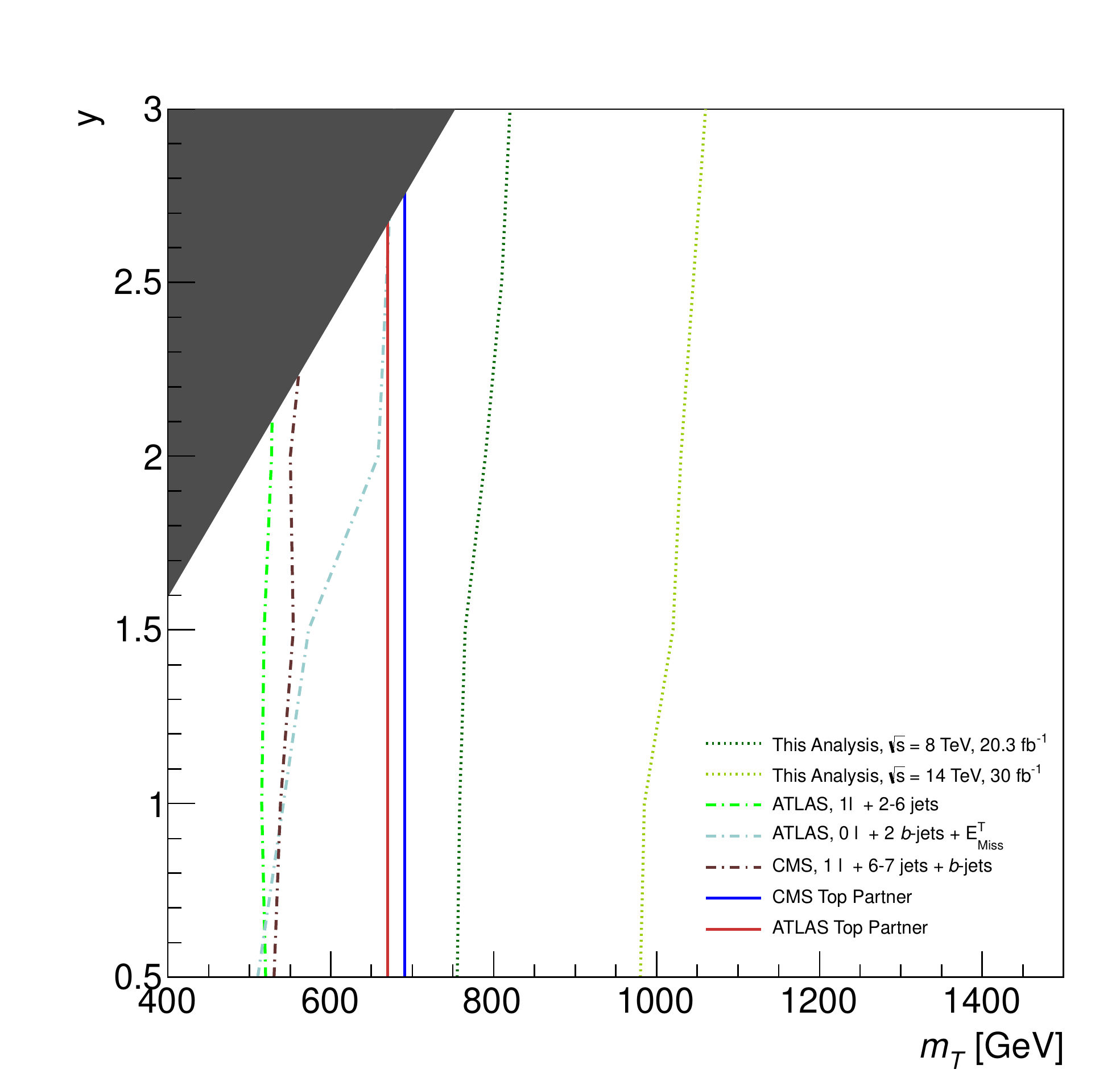}
\caption{Exclusion limits on the singlet model obtained from published analyses from the experiments (solid), from reinterpretations of analyses not explicitly interpreted in the context of top partners (dot-dashed), and projected exclusions with the analysis proposed here at $\sqrt{s} = 8$~\TeV~and  $\sqrt{s} = 14$~\TeV (dotted). In the dark grey region, the constraint from the known top quark mass cannot be satisfied.}
\label{fig:compHAllExclusionsSinglet}
\end{figure}

\begin{figure}[h!]
	\centering
	\includegraphics[width=0.8\linewidth]{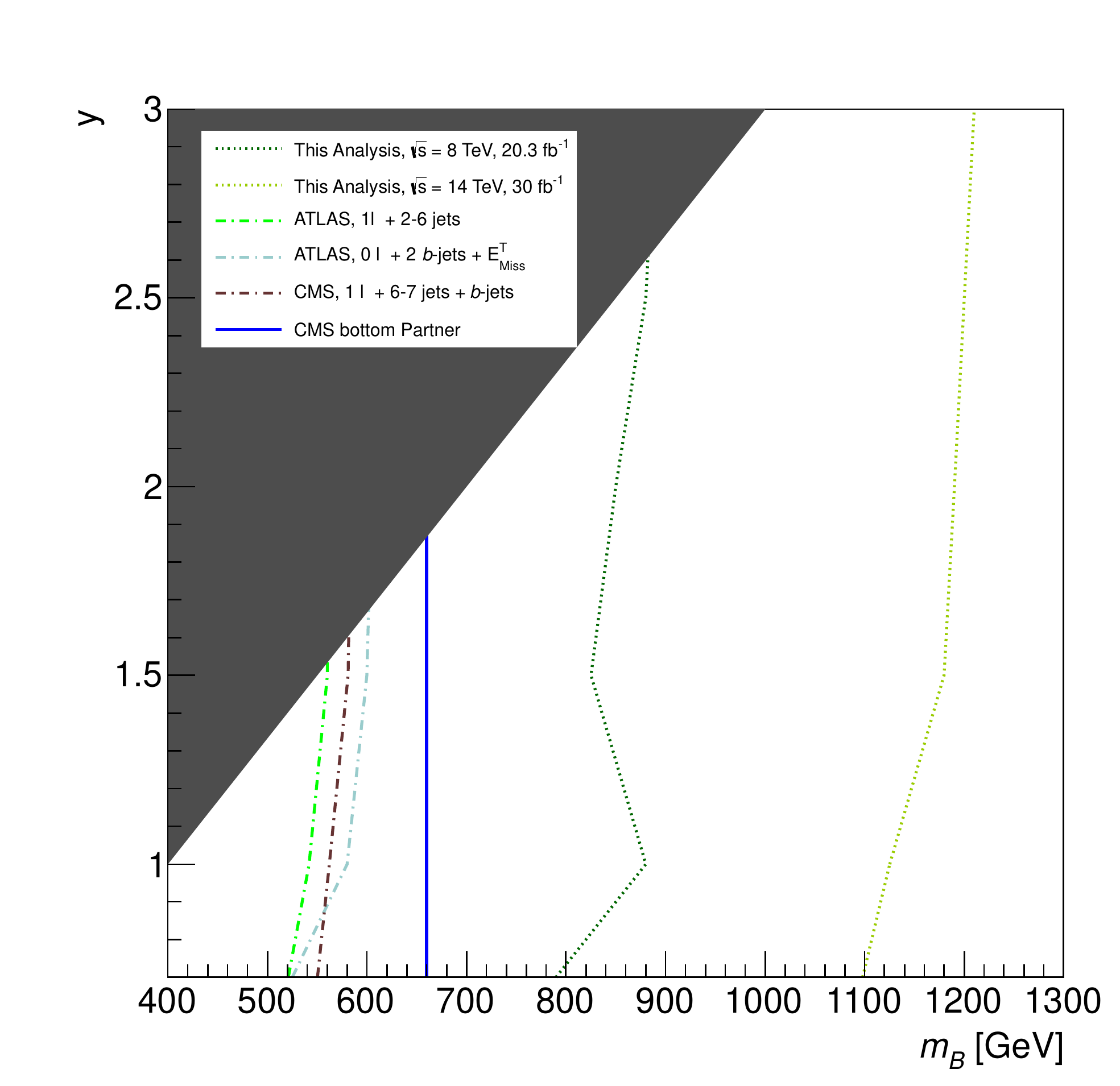}
\caption{Exclusion limits on the doublet model obtained from published analyses from the experiments (solid), from reinterpretations of analyses not explicitly interpreted in the context of top partners (dot-dashed), and projected exclusions with the analysis proposed here at $\sqrt{s} = 8$~\TeV~and  $\sqrt{s} = 14$~\TeV (dotted). In the dark grey region, the constraint from the known top and bottom quark masses cannot be satisfied.}
\label{fig:compHAllExclusionsDoublet}
\end{figure}


\section{Discussion}
\label{sec:disc}
We have presented a classification of the lightest fermionic
resonances, or top partners, in models in which the Higgs boson is a
composite resonance of strongly-coupled dynamics. The top partners are
expected to be reasonably light (and within the reach of the LHC), in
order to reproduce the measured Higgs boson mass. We find that there
are restrictions of the possible charges of top partners if we insist
either on a custodial symmetry to protect the rate for $Z
\rightarrow b\overline{b}$, or that the $t_R$ be wholly composite. In the first case, the top partner must transform in either the
$(2l+1,2l+1)^\pm$ or the $((2l+1,2l+3))$ irrep of the custodial
symmetry group $\custh$; in the second case, the top partner is
restricted to the $(2l+1,2l+1)$ irrep of $SU(2)^2$.

Using this classification, we have explored possible experimental
searches for top partners at the LHC. While top partners with
electric charges not equal to $\pm \frac{1}{3}, \pm \frac{2}{3},$ or
$\pm \frac{4}{3}$
should be relatively easy to find via excesses in events with
like-sign dileptons, other top partners, decaying to $t$ or $b$ quarks
and $W, H$ or $Z$ bosons with comparable branching fractions, may prove more difficult. For
these, we find that a single search that is based on a top quark veto,
but that it is otherwise reasonably inclusive, gives good sensitivity
throughout the space of models and parameters. Thus, it should be
relatively straightforward to either discover or exclude the composite
Higgs hypothesis at the LHC, by means of two generic searches.

\section*{Acknowledgements}
BG acknowledges
the support of the Science and Technology Facilities Council, the
Institute for Particle Physics Phenomenology, and King's College,
Cambridge and thanks R.~Contino and R.~Rattazzi for discussions. DS acknowledges the support of the Science and Technology Facilities Council, as well as Emmanuel College, Cambridge, and thanks O.Matsedonskyi for FeynRules help.
TM thanks C.~Lester for discussions on mass variables.

\clearpage

\appendix
\section{Group and Representation Theory of Top
  Partners \label{app:lie}}
In this Appendix, we discuss the necessary group and representation
theory for the classification of top partners  in a composite Higgs
model based on the coset $G/H$ given in \S \ref{sec:group}.
We begin by considering $H$ to be
locally isomorphic to $SU(2)^2$ and to be both connected
and simply-connected, {\em i.e.} $H \simeq SU(2)^2$. If $H$ contains $SU(2)^2$ as a proper subgroup, we can
make similar arguments, except that we should
replace $H$ by its $SU(2)^2$ subgroup and then pick out from the
representations thus obtained only those that can be lifted to reps of $H$. 
 Similarly, if $H$ is not simply-connected, then $H$ (or a subgroup thereof) can be
  obtained by taking the quotient of $ SU(2)^2$ with a non-trivial subgroup of its centre, $\mathbb{Z}_{2}^2$.
These quotients are isomorphic to $SU(2) \times
SO(3)$, $SO(3)^2$, and $SO(4)$ (which is obtained by taking
the quotient with the diagonal $\mathbb{Z}_2$ subgroup of $\mathbb{Z}_{2}^2$).\footnote{These results are easily established by means of the
following explicit homomorphisms: (i) Consider $(x,y,z) \in
\mathbb{R}^3$ and let $M \equiv \begin{pmatrix} z & x+iy \\ x-iy &
  z \end{pmatrix}$. Then $U \in SU(2)$ acting as $M \mapsto UMU^*$
effects an orthogonal transformation of determinant $+1$ on
$\mathbb{R}^3$ with kernel $U \in \{I,-I\}$; (ii) Consider $(x,y,z,w) \in
\mathbb{R}^4$ and let $M \equiv \begin{pmatrix} z+iw & x+iy \\ -x+iy &
  z-iw \end{pmatrix}$. Then $(U,V) \in SU(2) \times SU(2)$ acting as
$M \mapsto VMU^{-1}$ preserves $\mathrm{det} M = x^2 + y^2 + z^2 +
w^2$ and so is an orthogonal transformation on
$\mathbb{R}^4$ with kernel $(U,V) \in \{(I,I),(-I,-I)\}$. Direct
computation establishes that the map has determinant +1.} 
The only change in the results obtained below is that the possible
irreps for the top partner are limited to those
irreps of the covering group $SU(2)^2$ that restrict to the identity on
the subgroup used to form the quotient, namely those $(2l+1,2r+1)$ with even values of
$2r$, $2l$ and $2r$, and $2(l+r)$, respectively.

If we wish to protect the rate for $Z\rightarrow b\overline{b}$,
then it is useful to also consider the case where $H$ is a
(disconnected) semi-direct product of $SU(2)^2$ with $\mathbb{Z}_2$.
Recall that given two groups, $N$ and $K$, and a homomorphism
$\phi: K \rightarrow \mathrm{Aut} (N)$, the semi-direct product group
$N \rtimes K$ is the set $N \times K$ together with the multiplication
operation defined by $(n_1,k_1) . (n_2,k_2) \equiv ( n_1 \phi_{k_1}
(n_2),k_1 k_2)$. Here, we let $N = SU(2)^2$ and $K = \mathbb{Z}_2$, with $\phi$ mapping $\mathbb{Z}_2$ to the (outer) automorphisms of $N$ obtained by permuting the two
$SU(2)$s.\footnote{In \cite{Agashe:2006at}, this group is written as $SU(2) \otimes
  SU(2) \otimes P_{LR}$.}

The irreps of $\custh$ can be obtained mechanically by
inducing them from the irreps of the subgroups $SU(2)^2$ and
$\mathbb{Z}_2$, but it is
easy enough to guess the irreps directly and then prove that they are
all irreps. The result is that there are three classes of irreps.
Two of these classes, which we
denote by $(2l+1,2l+1)^{\pm}$, are equivalent to
\begin{align} 
\label{eq:pmrep}
 D^\pm (g_L,g_R,+e) &= D^l_{\alpha \beta} (g_L)
 D^l_{\dot{\alpha} \dot{\beta}}(g_R) \nonumber \\
 D^\pm (g_L,g_R,-e) &=\pm D^l_{\alpha \dot{\beta}}(g_L)
 D^l_{\dot{\alpha} \beta}(g_R)
\end{align}
where $D^l(g)$ is any matrix representing the $SU(2)$ element $g$ in
the irrep $2l+1$. When restricted to $SU(2)^2$, both of these irreps
reduce to the direct product irrep, but they are inequivalent as irreps
of $\custh$. 
The third class of irreps, which we denote by
$((2l+1,2r+1))$ with $l\neq r$, are given by
\begin{align} 
D(g_L,g_R,+e) = \left( \begin{matrix} D^l( g_L) \otimes D^r(g_R) & 0 \\ 0 & D^l(g_R) \otimes D^r(g_L) \end{matrix} \right) \nonumber \\
D (g_L,g_R,-e) = \left( \begin{matrix} 0 & D^l(g_L) \otimes D^r(g_R) \\ D^l(g_R) \otimes D^r(g_L) & 0 \end{matrix} \right) .
\label{eq:sumrep}
\end{align}
Their restriction to
 $SU(2)^2$ is a direct sum, $(2l+1,2r+1)
 \oplus (2r+1,2l+1)$.  

We prove that (\ref{eq:pmrep}-\ref{eq:sumrep}) are all irreps by standard methods in the
representation theory of compact Lie groups \cite{Brocker}, showing
that the characters of the irreps form a complete, orthonormal set of
functions on the conjugacy classes of the group. 

We warm up with $SU(2)$, where every element is conjugate (denoted by $\sim$) to an element
of the form $\mathrm{diag} (e^{it} ,
  e^{-it} )$. Now $t \sim t +2\pi$ and moreover $t \sim -t$, as can be
  shown by conjugating
with $\begin{pmatrix} 0 & 1 \\ -1 &
  0 \end{pmatrix} \in SU(2)$.
So the
class functions are even, periodic
functions on $\mathbb{R}$. Now, the usual irrep with multiplicity $2j+1$ has
class representative $\mathrm{diag} ( e^{2ijt} ,  e^{2i(j-1)t}, \dots  ,e^{-2ijt} )$
and character $\chi^{j} (t) = \frac{\sin (2j+1)t}{\sin t} = \cos 2jt +
\chi^{j-\frac{1}{2}}\cos t $, so it is clear that by taking linear
combinations with different $j$, we may obtain the complete set of
even, periodic functions $1, \cos t, \cos 2t, \dots $ exactly once.
Finally, the normalized group-invariant measure on $SU(2)$ is the same as the one
obtained from the round
metric on $S^3$, {viz.} $\frac{1}{2\pi^2} 
\int_{0}^{\pi} \sin^2 t dt \int_0^\pi \sin \theta d \theta \int_{0}^{2\pi} d \phi $, which
reduces to $\frac{2}{\pi} \int_0^\pi dt \sin^2t $ on class functions.
Since the characters of the above reps of $SU(2)$ satisfy
\begin{gather}
\frac{2}{\pi} \int_0^\pi dt \sin^2t \; \chi^{j*} (t) \chi^k (t) =\delta^{jk},
\end{gather} 
they are orthonormal and so they are indeed a complete set of irreps.

For $SU(2)^2$, the
classes may be represented
by $\left( \mathrm{diag} (e^{it_L} ,
    e^{-it_L} ) , \mathrm{diag} (e^{it_R} ,
    e^{-it_R} )\right) $ and the character of rep $(2l+1,2r+1)$ is
  then
the product of the corresponding $SU(2)$ characters and
  the measure is the product of the two $SU(2)$ measures.
It is then immediate that these are inequivalent irreps and that they are all of the irreps.

For $SU(2)^2 \rtimes \mathbb{Z}_2$, $(g_L,g_R,e)
\sim (g_R,g_L,e)$ and so the classes in the component
connected to the identity may be represented by
$\left( \mathrm{diag} (e^{it_L} ,
    e^{-it_L} ) , \mathrm{diag} (e^{it_R} ,
    e^{-it_R} ),e \right) $, where we restrict class functions to be even, periodic, and symmetric
under the interchange $t_L \leftrightarrow t_R$. In the component that
is not connected to the identity, we find that conjugating
with $(h_L,h_R,e)$ and $(h_L,h_R,-e)$ yields
$(g_L,g_R,-e) \sim (h_L^{-1} g_L h_R,h_R^{-1} g_R h_L,-e)$ and $(g_L,g_R,-e) \sim (h_R^{-1} g_R
h_L,h_L^{-1} g_L h_R,-e)$. The second of these
conjugations with $h_L =g_L$ and $h_R =e$ implies that
$(g_L,g_R,-e) \sim (k,e,-e)$, where $k = g_R g_L$, and the first conjugation with $h_L =h_R =h$ then implies that
$(k,e,-e) \sim (h^{-1} k h, e,-e) $. The upshot is that classes in the
component disconnected from the identity can be parameterised
similarly to those of $SU(2)$, {\em viz.} as $\left( \mathrm{diag} (e^{it}, 
    e^{-it} ) , e,-e \right) $, with class functions
  that are even and periodic in $t$.

The characters of the 
$(2l+1,2l+1)^\pm$  reps are given by
\begin{align} \label{eq:char1}
\chi^\pm (+e) &= \frac{\sin (2l+1)t_L}{\sin t_L}\frac{\sin (2l+1)t_R}{\sin
  t_R} \\
\chi^\pm (-e) &= \pm \frac{\sin (2l+1)t}{\sin t}
\end{align}
and the characters of the 
$((2l+1,2r+1))$  rep are given by
\begin{align} \label{eq:char2}
\chi (+e) &= \frac{\sin (2l+1)t_L}{\sin t_L}\frac{\sin (2r+1)t_R}{\sin
  t_R} + (L \leftrightarrow R) \\
\chi (-e) &= 0,
\end{align}
where $\chi (+e)$ and $\chi (-e)$ are the characters of the components connected to and disconnected from the identity.

As a manifold, the group is just $SU(2)^2 \times
\mathbb{Z}_2$, and so the normalized group-invariant measure is just
one-half of that for $SU(2)^2$ on each connected component. On the
component connected to the identity, this reduces to 
\begin{gather}
\frac{2}{\pi^2} \int_0^\pi dt_L \sin^2t_L \int_0^\pi dt_R
\sin^2t_R 
\end{gather}
on class functions, while on the component disconnected from the
identity it reduces to
\begin{gather}
\frac{1}{\pi} \int_0^\pi dt \sin^2t.
\end{gather}
One may then easily check that the characters (\ref{eq:char1}-\ref{eq:char2}) are orthonormal
with respect to the measure. They are, moreover, a complete set of functions with the given
properties.
Note that the characters furnish a basis for independent functions on the two disconnected components, as expected.

So far, we have established that the top partners can come in reps
formed from the irreps $(2l+1,2r+1)$ (without custody of $Z
\rightarrow b\overline{b}$) or  $(2l+1,2l+1)_\pm$ and $((2l+1,2r+1)) $ (with custody of $Z
\rightarrow b\overline{b}$). To establish further constraints on the
reps, we need to consider the non-linearly realized group, $G$, and
its irreps.

The minimal choice for $G$ (in order to furnish a SM Higgs doublet \cite{Agashe:2004rs}) is a group locally isomorphic to
$SO(5)$. By similar arguments to those given above for $H$, it
suffices to consider the universal cover, 
{\em viz.} $Sp(2)$. We define $Sp(n)$ as the group of unitary transformations on
  $n$-dimensional quaternions, $\mathbb{H}^n$, preserving the bilinear
  $\overline{x} \cdot y$, where we define conjugation of quaternions
  by $\overline{x_1 + ix_2 +jx_3 +k x_4} \equiv x_1 - ix_2 - jx_3 - k
  x_4$. Thus $Sp(1)$ is isomorphic to $SU(2)$, and since $Sp(2) \supset
Sp(1)^2$, we have that $Sp(2) \supset
SU(2)^2$. Moreover, the outer automorphism of $SU(2)^2$ that permutes the two $SU(2)$s is an inner automorphism of
$Sp(2)$, so $Sp(2) \supset
\custh$. Explicitly, the embeddings for the
components connected to, and disconnected from, the identity are
$\begin{pmatrix} g_L & 0\\ 0 &
  g_R \end{pmatrix}$ and $\begin{pmatrix} 0 & g_L\\ g_R &
  0 \end{pmatrix}$, respectively. 

We next need to know the irreps of $Sp(2)$ and how they
restrict to the subgroup $\custh$ (and its subgroup $SU(2)^2$, for
models without custodial protection of $Z \rightarrow b
\overline{b}$). At least for the subgroup $SU(2)^2$, all this can be found
in the literature \cite{Kemmer}. An irrep of $Sp(2)$ can be labelled
by $(m,k)$, where $m \geq k \geq 0$ and $m$ and $k$ are either both
integer or both half-integer, and has dimension $ (2k + 1)(2m + 3)(m +
k + 2)(m- k + 1)/6$. Under restriction to $SU(2)^2$, the irreps $(2l+1,2r+1)$ of
$SU(2)^2$ arise with multiplicity at most one (since $Sp(2)/SU(2)^2$
is a symmetric space \cite{Helgason}), and are given by half-integers
$ 0 \leq l,r \leq (k+m)/2$, with the restriction that for a fixed
value of $l$, $r \in \{|k-l|, |k-l|+1, \dots, \mathrm{min} (m-l,k+l)\}$
(and {\em vice versa}) for $l \leftrightarrow r$. For $\custh$, we
need only figure out whether the $(2l+1,2l+1)$ irreps that appear
correspond to $(2l+1,2l+1)^+$ or $(2l+1,2l+1)^-$. This can be done by
constructing the relevant $Sp(2)$ irreps
as tensor products of the fundamental.
Indeed, the $(\frac{1}{2},\frac{1}{2} )$ irrep has characters
\begin{align}
\chi^4(+e) &= \cos t_L + \cos t_R, \\
\chi^4(-e) &= 0,
\end{align}
so the antisymmetric and symmetric parts of its direct product with
itself have characters
given by $\frac{1}{2} (\mathrm{tr}^2 D \mp \mathrm{tr} D^2)$, or
\begin{align}
\chi^{[4 \times 4]} (+e) &= 1 + 1 + \frac{\sin 2t_L }{\sin
  t_L}\frac{\sin 2t_R }{\sin t_R} \\
\chi^{[4 \times 4]}(-e) &= 1 - 1 - 2\cos t,
\end{align}
and
\begin{align}
\chi^{(4 \times 4)}(+e) &= 4\cos^2 t_L  + 4\cos^2 t_R-2 + 4 \cos t_L \cos
t_R , \nonumber \\
&= \frac{\sin 3t_L }{\sin t_L} + \frac{\sin 3t_R }{\sin t_R} + \frac{\sin 2t_L }{\sin t_L}\frac{\sin 2t_R }{\sin t_R}\\ 
\chi^{(4 \times 4)}(-e) &= 0+ 2\cos t, 
\end{align}
respectively. The first of these shows that $(1,0) \rightarrow (1,1)^- + (2,2)^-$ (and, of course $(0,0)
\rightarrow (1,1)^+$) and the second shows that $(1,1) \rightarrow ((1,3)) \oplus (2,2)^+$.
 In general ,we find that the irrep $(m,k)$ restricts to irreps
$(2l+1,2l+1)^\pm$ with $\pm$ given by $(-1)^{(m+k)}$.

These results have four immediate corollaries that that are useful for the
classification of top partners: (i) the irrep $(2,2r+1)$ of $SU(2)^2$
is contained in all irreps with $k \leq r+\frac{1}{2}$ and $m$
sufficiently large, where $k$ (and $m$) must
be integer or half-integer if $r$ is integer or half-integer,
respectively; (ii) for $m$ sufficiently large, these irreps branch to
the $SU(2)^2$
irreps $(2l+1,2r^\prime+1)$, with $r^\prime \in \{ |l-k|, |l-k|+1,
\dots l+k \}$; (iii) there are no additional possibilities at small
$m$; (iv) only the $(m,0)$ irreps of $Sp(2)$ contain
singlets of $SU(2)^2$.

Using these results, we see that the possible irreps for top
partners depend on whether we wish to have
custodial protection of $Z \rightarrow b \overline{b}$
and whether we wish the $t_R$ to be fully composite. Custodial
protection of $Z \rightarrow b \overline{b}$ (or $Z \rightarrow b_L
\overline{b_L}$, to be precise)
is achieved \cite{Agashe:2006at} by insisting that the Lagrangian describing the
 composite sector and the elementary $b_L$ be invariant under the subgroup $U(1)^2\rtimes \mathbb{Z}_2$. After electroweak symmetry breaking, only the $U(1)
\times \mathbb{Z}_2$ (note that it is now a direct product) subgroup
is linearly realized, where the $U(1)$ is the diagonal combination of the
2 original $U(1)$s. The irreps of $U(1)^2\rtimes \mathbb{Z}_2$ are, in
a similar notation to that used for $\custh$, $(q,q)^\pm$ (of
dimension 1) and $((q_L,q_R))$ (of dimension 2), where $q,q_L,$ and
$q_R$ are the charge $q,q_L,$ and
$q_R$ irreps of $U(1)$. The branching rules under $U(1)^2\rtimes
\mathbb{Z}_2 \rightarrow U(1)
\times \mathbb{Z}_2$ are $(q,q)^\pm \rightarrow q$ and $((q_L,q_R))
\rightarrow (q_L+q_R) \oplus (q_L+q_R)$. Since $U(1)$ is an unbroken
symmetry,
$q$ and $q_L + q_R$ are conserved, and so if the $b_L$ carries the
$(q,q)^\pm$ irrep, its coupling to the $Z$, which is a linear
combination of $q$ and the conserved electric charge, is
protected. But $b_L$ must also transform as a doublet of $SU(2)_L$ and
so the only suitable irreps of $\custh$ are the $(2,2)^\pm$.

Now, a top partner in irrep $\rho_H$ of $H$ in such a model must be
able to mix with the elementary $Q_L$ via the coset representative,
and the usual sigma model lore tells us that this can happen
only if the $Q_L$ is assigned to a spurion in a representation $\rho_G$ of $G$
that contains $\rho_H$ on restriction to $H$. 

If we want the $t_R$ to be fully composite, it must come in a 1-d
irrep of $H$, {\em viz.} $(1,1)^\pm$. For if it does not, then there will be
additional massless states before EWSB. This too can mix with the
$Q_L$
via the coset representative only if $\rho_G$ contains a $(1,1)^{(\pm)}$. 

 Thus, if, on the one hand, we desire
full $t_R$ compositeness, then the top partner must come in an irrep
of $H$ that is contained in an irrep of $G$ that contains a
$(1,1)^{\pm}$. The only such irreps of $G$ are of the
form $(m,0)$, with $m \in \mathbb{N}$ and these contain exclusively the irreps
$(2l+1,2l+1)^{(-1)^m}$ with integer $ 2l \leq m$. These irreps also
all contain a $(2,2)^{\pm}$, so can feature protection of $Z\rightarrow b\overline{b}$.

If, on the other hand, we desire only protection of $Z\rightarrow
b\overline{b}$, then $\rho_G$ need only contain
a $(2,2)^\pm$  and we find in addition the $G$ irreps $(m,1)$
with $m \in \mathbb{N}$. These contain the $H$ irreps $(2l+1,2l+1)^{(-1)^{m+1}}$ with $1 \leq 2l \leq m$, together with $((2l+1,2l+3))$ for $0 \leq 2l \leq m$.

Finally, if we desire neither $t_R$ compositeness nor protection of $Z\rightarrow
b\overline{b}$, we need only that the rep $\rho_G$ contain a $(2,2r^\prime+1)$ of
$SU(2)^2$, so as to contain the $Q_L$. For a suitable $r^\prime$, it is then possible to put a
top partner in any irrep $(2l+1,2r+1)$. 
\section{Details of minimal model without $Z\rightarrow b\overline{b}$ protection}
\label{app:model}
We define the generators in the $(\frac{1}{2},\frac{1}{2})$ rep of
$Sp(2)$ as
\begin{align*}
 T_L^i = \frac{1}{\sqrt{2}} \left( \begin{matrix} \sigma^i & 0 \\ 0 & 0 \end{matrix} \right) \;
 T_R^i = \frac{1}{\sqrt{2}} \left( \begin{matrix} 0 & 0 \\ 0 & \sigma^i \end{matrix} \right) \;
 T^\alpha = \frac{1}{2} \left[ \left( \begin{matrix} 0 & \sigma^i \\ \sigma^i & 0 \end{matrix} \right) , \left( \begin{matrix} 0 & -i 1 \\ i 1 & 0 \end{matrix} \right) \right]
\end{align*}
where $T_L^i,T_R^i$ ($i=1,2,3$) are the generators of the $SU(2)^2$ subgroup. The $T^\alpha$ ($\alpha=1,2,3,4$) may be taken in combination with four real fields to form the coset representative
\begin{equation}
 U = \exp \left( \frac{2 i  \Pi^\alpha T^\alpha}{f} \right) = \left( \begin{matrix} 1_{2\times 2} \cos \epsilon & 1_{2\times 2} \sin \epsilon \\ -1_{2\times 2} \sin \epsilon & 1_{2\times 2} \cos \epsilon \end{matrix} \right)
\end{equation}
where we identify $\Pi^{1,2,3}$ as the Goldstone bosons and $\Pi^4$ as the Higgs field of the SM (in the limit $f \rightarrow \infty$). Thus the last equality gives $U$ in unitary gauge, where $\epsilon \equiv \frac{\langle \Pi^4 \rangle}{f}$.

We define the elementary Weyl fermions of the model as
\begin{equation}
 Q_L = \left( \begin{matrix} t_L & b_L & 0 & 0 \end{matrix} \right)^T,\;  t_R = \left( \begin{matrix} 0 & 0 & t_R & 0 \end{matrix} \right)^T,\; b_R = \left( \begin{matrix} 0 & 0 & 0 & b_R \end{matrix} \right)^T
\end{equation}
and for the top partner
\begin{equation}
 \Psi = \left( \begin{matrix} 0 & 0 & T & B \end{matrix} \right)^T. 
\end{equation}

In order to construct Lagrangian terms involving the $\Pi$ fields that are invariant under $Sp(2)$, we consider the quantities
\begin{align}
 \mathcal{A}_\mu = \frac{1}{\sqrt{2}} (g W^1 T^1_L + g W^2 T^2_L + g W^3 T^3_L + g^\prime B T^3_R) \\
 U^\dagger (\mathcal{A}_\mu + i \partial_\mu) U = -d^i_\mu T^i - e^i_{L \mu} T^i_L - e^i_{R \mu} T^i_R .
\end{align}
One can show that the $e$ terms transform as a gauge connection under $SU(2)\times U(1)$, and may thus form part of the covariant derivative of $\Psi$. The $d$ terms transform linearly under $SU(2)^2$, are comprised of the derivatives of the $\Pi$ fields, and make up the Higgs kinetic terms.

All together, the Lagrangian is
\begin{align}
 \mathcal{L} &= i \overline{Q_L} \slashed{D} Q_L + i \overline{t_R} \slashed{D} t_R + i \overline{b_R} \slashed{D} b_R + \overline{\Psi} (i \slashed{D} - \slashed{e} - M) \Psi + \frac{f^2}{8} d^i_\mu d^\mu_i \nonumber \\
 & -y f \overline{Q_L} U \Psi_R - y_R^t f \overline{t_R} U \Psi_L - y_R^b f \overline{b_R} U \Psi_L + \mathrm{h.c}
\end{align}
The covariant derivative of the elementary fields is that of the
Standard Model, whereas only colour and part of the hypercharge
interactions are present in $D_\mu \Psi = \partial_\mu \Psi - \frac{1}{6} i g^\prime B_\mu 1_{4\times 4} - i g_S G_\mu$.\footnote{In order to obtain the correct electric charges
  for the SM fermions, an extra symmetry $U(1)_X$ is included in both $G$ and $H$, making the linearly realized symmetry group $U(1)_X \times
  SU(3)_c \times SU(2)_L \times SU(2)_R (\rtimes \mathbb{Z}_2)$. We
  then gauge a subgroup of this, namely $SU(3)_c \times SU(2)_L \times
  U(1)_Y$, such that the hypercharge is $Y=T^3_R+X$. Thus the gauge connections pertaining to colour and {\em part} of $U(1)_Y$ appear in $D$, whereas weak isospin and the other part of $U(1)_Y$, these being a subgroup of the $SU(2)^2$ considered above, are present
  in $e$.}

To obtain the correct $W$ and $Z$ masses after EWSB, $\epsilon$ is fixed by $v = f \sin \epsilon \cos \epsilon = 246 \Gev$.

The Yukawa interactions mix the elementary top and bottom with the $T$
and $B$. Before EWSB ($\langle \Pi^4 \rangle = 0$), the top and bottom
remain massless, whereas after ($\langle \Pi^4 \rangle \neq 0$) the $t,b,T$ and $B$ masses are functions of $y,y^t_R,y^b_R,M$ and $f$. Two of these must be fixed to obtain the correct top and bottom masses.
\bibliographystyle{JHEP}
\bibliography{top_partners}

\end{document}